\documentclass[aps,prl,10pt,amsmath,amssymb,twocolumn,superscriptaddress]{revtex4-2}
\usepackage{natbib}
\usepackage{graphicx}         % Include figure files
\usepackage{dcolumn}          % Align table columns on decimal point
\usepackage{bm}               % bold math
\usepackage{upgreek}
\usepackage{booktabs}
\usepackage{gensymb}
\usepackage{array}
\usepackage[colorlinks=true,urlcolor=blue,breaklinks=true]{hyperref}
\usepackage{xcolor}
\usepackage{chemformula}

\begin{document}

\title{Quantum transport in topological surface states of Bi$_2$Te$_3$ nanoribbons}

\author{D. Rosenbach}
\affiliation{Peter Gr\"unberg Institut (PGI-9) and JARA-Fundamentals of Future Information Technology, J\"ulich-Aachen Research Alliance, Forschungszentrum J\"ulich, 52425 J\"ulich, Germany}

\author{N. Oellers}
\affiliation{Peter Gr\"unberg Institut (PGI-9) and JARA-Fundamentals of Future Information Technology, J\"ulich-Aachen Research Alliance, Forschungszentrum J\"ulich, 52425 J\"ulich, Germany}

\author{A.R. Jalil}
\affiliation{Peter Gr\"unberg Institut (PGI-9) and JARA-Fundamentals of Future Information Technology, J\"ulich-Aachen Research Alliance, Forschungszentrum J\"ulich, 52425 J\"ulich, Germany}

\author{M. Mikulics}
\affiliation{Ernst Ruska-Centre for Microscopy and Spectroscopy with Electrons, Materials Science and Technology (ER-C-2), Forschungszentrum J\"ulich, 52425 J\"ulich, Germany}

\author{J. K\"olzer}
\affiliation{Peter Gr\"unberg Institut (PGI-9) and JARA-Fundamentals of Future Information Technology, J\"ulich-Aachen Research Alliance, Forschungszentrum J\"ulich, 52425 J\"ulich, Germany}

\author{E. Zimmermann}
\affiliation{Peter Gr\"unberg Institut (PGI-9) and JARA-Fundamentals of Future Information Technology, J\"ulich-Aachen Research Alliance, Forschungszentrum J\"ulich, 52425 J\"ulich, Germany}

\author{G. Mussler}
\affiliation{Peter Gr\"unberg Institut (PGI-9) and JARA-Fundamentals of Future Information Technology, J\"ulich-Aachen Research Alliance, Forschungszentrum J\"ulich, 52425 J\"ulich, Germany}

\author{S. Bunte}
\affiliation{Helmholtz Nanoelectronic Facility (HNF), Forschungszentrum J\"ulich, 52425 J\"ulich, Germany}

\author{D. Gr\"utzmacher}
\affiliation{Peter Gr\"unberg Institut (PGI-9) and JARA-Fundamentals of Future Information Technology, J\"ulich-Aachen Research Alliance, Forschungszentrum J\"ulich, 52425 J\"ulich, Germany}

\author{H. L\"uth}
\affiliation{Peter Gr\"unberg Institut (PGI-9) and JARA-Fundamentals of Future Information Technology, J\"ulich-Aachen Research Alliance, Forschungszentrum J\"ulich, 52425 J\"ulich, Germany}

\author{Th. Sch\"apers}
\email{th.schaepers@fz-juelich.de}
\affiliation{Peter Gr\"unberg Institut (PGI-9) and JARA-Fundamentals of Future Information Technology, J\"ulich-Aachen Research Alliance, Forschungszentrum J\"ulich, 52425 J\"ulich, Germany}

\hyphenation{}
\date{\today}

\begin{abstract}
Quasi-1D nanowires of topological insulators are emerging candidate structures in superconductor hybrid architectures for the realization of Majorana fermion based quantum computation schemes. It is however technically difficult to both fabricate as well as identify the 1D limit of topological insulator nanowires. Here, we investigated selectively-grown Bi$_2$Te$_3$ topological insulator nanoribbons and nano Hall bars at cryogenic temperatures for their topological properties. The Hall bars are defined in deep-etched Si$_3$N$_4$/SiO$_2$ nano-trenches on a silicon (111) substrate followed by a selective area growth process via molecular beam epitaxy. The selective area growth is beneficial to the device quality, as no subsequent fabrication needs to be performed to shape the nanoribbons. Transmission line measurements are performed to evaluate contact resistances of Ti/Au contacts applied as well as the specific resistance of the Bi$_2$Te$_3$ binary topological insulator. In the diffusive transport regime of these unintentionally $n$-doped Bi$_2$Te$_3$ topological insulator nano Hall bars, we identify distinguishable electron trajectories by analyzing angle-dependent universal conductance fluctuation spectra. When the sample is tilted from a perpendicular to a parallel magnetic field orientation, these high frequent universal conductance fluctuations merge with low frequent Aharonov--Bohm type oscillations originating from the topologically protected surface states encircling the nanoribbon cross section. For 500\,nm wide Hall bars we also identify low frequent Shubnikov--de Haas oscillations in the perpendicular field orientation, that reveal a topological high-mobility 2D transport channel, partially decoupled from the bulk of the material.
\end{abstract}
\maketitle

\section{I. Introduction}
Three-dimensional topological insulators (3D TIs) are a new class of materials that have a bulk electronic gap but highly conductive surface states, which promise a gapless, Dirac-like dispersion relation and spin-momentum locking of charge carriers occupying these surface states \cite{Fu2007,Ando13}. 3D TIs  are no longer only interesting for basic research but this new material class has slowly matured as candidates for a wide spectrum of applications, including the possible use for topology based quantum computation schemes \cite{Fu2008,Manousakis2017,Hyart13}. In topological qubits the quantum state is decoded via the spatial arrangement of two Majorana zero modes (MZMs) \cite{Nayak2008}. These arise e.g. in the vortex core of a type-II, $s$-wave superconductor at the interface towards a 3D TI \cite{Sun2016}. Another possibility is to design quasi-1D nanowires of 3D TIs, proximitized by an $s$-wave superconductor, where two MZMs will arise at both ends of the nanowire \cite{Cook2011,Juan2014}. In this context it is highly important to have a basic understanding of quantum transport in these TI nanowires.\\
\indent Conventional 3D TIs are binary compounds such as Bi$_2$Te$_3$ \cite{Krumrain2011},  Bi$_2$Se$_3$ \cite{Benia2011,Zhang2011a},  Sb$_2$Te$_3$ \cite{Takagaki2012}, as well as alloys of these elements \cite{Zhang2011,Ren2011}. The observation of periodic Aharonov--Bohm (AB) oscillations in an external parallel magnetic field has previously been reported to indicate the existence of highly coherent two-dimensional sheets on the perimeter of such topological insulator nanowires \cite{Peng2010,Zhang2010,Xiu2011,Cho2015,Jauregui2015,Arango16}. For a quasi 1D nanowire, due to the inclusion of the Berry-phase of a particle traversing the perimeter, the band structure is determined to be gapped \cite{Zhang2009,Bardarson2010,Bardarson2013}. By applying a magnetic flux parallel to the nanowire axis $\Phi = \pm \Phi_0/2$, where $\Phi_0 = h/e$, this gap will be closed. The non-degenerate, topologically protected linear surface bands will re-emerge periodically with a period of a full flux quantum $\Phi = (n+1/2)\Phi_0$ threading the wire. In a perfectly ballistic quasi-1D wire the magnetoconductance is therefore expected to oscillate with an amplitude of $\Delta G=\pm e^2/h$, due to the periodic inclusion of the topologically protected linear surface bands.\\
\indent Making use of a selective area growth approach, nanowires of aforementioned van der Waals materials can as well be deposited by molecular beam epitaxy (MBE) \cite{Moors2018, Schueffelgen2019, Weyrich2017, Koelzer2019}. Due to the layer growth inside predefined nanotrenches, these nanowires have a rectangular cross-section and are therefore regarded to as nanoribbons. Compared to the cylindrical nanowire geometry different matching conditions need to be chosen for the determination of the surface bands \cite{Brey2014} as four individual surfaces along the perimeter need to be considered. As a scalable top-down approach, these selectively grown nanoribbons are beneficial for desired Majorana surface architectures \cite{Bocquillon2019}. However, MBE grown 3D TI compounds usually suffer from a high unintentional background doping \cite{Scanlon2012,Weyrich2017}. The reasons for this are local defects in the crystal lattice such as antisite defects and vacancies during thin film deposition. Therefore, it is difficult to characterize the surface state properties of these materials electrically at low temperatures, as these are usually superimposed by bulk contributions \cite{Weyrich2017}. In such disordered nanoribbons with non-negligible bulk contributions $G \gg e^2/h$, the expected amplitude for the flux periodic oscillations in the nanoribbons cross section deviates from the simple periodic inclusion of one additional transport channel \cite{Bardarson2010,Bardarson2013}. The magnetoconductance along these diffusive nanoribbons is enhanced, whenever time-reversal symmetry is established. The reason therefore is the destructive interference in time-reversed loops, the weak-antilocalization (WAL) effect \cite{Hikami1980}. Time-reversal symmetry is restored at every integer and half-integer value of the magnetic flux quantum $\Phi = (n\cdot 1/2) \Phi_0$ \cite{Bardarson2010}. In bulk systems that exhibit strong spin-orbit coupling, the WAL effect is found to dominate the magnetoconductance oscillations \cite{Bergmann1982}. This makes it additionally difficult to deem AB-type oscillation features within the magnetoconductance of a nanoribbon to originate from topologically protected surface states. The transition in between rather bulk related, diffusive quantum interference modulations to more quasi-ballistic surface related quantum transport has just recently been reported in gate-dependent measurements on etched HgTe nanoribbons \cite{Ziegler2018}.\\
\indent Here, Bi$_2$Te$_3$ nanoribbons and nano-Hall bars have selectively been grown and electrically characterized at cryogenic temperatures. The channel width ranges from 50\,nm up to 500\,nm. Nanotrenches are defined in a Si$_3$N$_4$/SiO$_2$ layer stack on top of a Si(111) substrate. During growth the Bi$_2$Te$_3$ is selctively grown only within these nanotrenches, forming 3D TI nanoribbons of an approximately rectangular cross section. Furthermore, in order to protect the pristine topological surface states, a 5\,nm thin Al$_2$O$_3$ capping layer is deposited \textit{in situ}. A hard-capping such as this has previously been reported to effectively protect the topological surface states from degradation \cite{Lang2012,Schueffelgen2019}, that is caused due to oxidation and adhesion of water.\\
\indent Previous studies on MBE grown 3D TI nanoribbons have shown that magnetotransport measurements reveal favored, defect-based electron trajectories ('fingerprints') in these TI nanodevices, that mainly originate from 2D planes, parallel to the sample surface \cite{Koelzer2019}. These are dependent on the individual arrangement of scattering centers in the bulk of the nanoscale device, and therefore unique. One of the prerequisites to observe such quantum mechanical modulations to the mesoscopic magnetoconductance is that the length of trajectories under consideration stay within the limit of the phase-coherence length $l \leq l_{\phi}$, which is different for the bulk and for the surface state charge carriers \cite{Dufouleur2017}.\\
\indent Electrical characterization at cryogenic temperatures includes two-terminal measurements of selectively grown micro- and nanoribbons using the transmission line method (TLM). Furthermore, a detailed analysis of magnetoconductance modulations of nano-TLM and nano Hall bar devices has been performed. The magnetic field orientation was changed from an in-plane parallel orientation to a perpendicular out-of-plane orientation. In the perpendicular field orientation traceable universal conductance fluctuation (UCF) spectra can be identified. Performing Fast Fourier transformation and correlation field analysis on the magnetoconductance data the bulk phase coherence length has been determined. In the parallel field orientation, AB interference modulations can be identified that are restricted to the cross section of the nanoribbons. The AB phase can also be observed for nanoribbons with a perimeter, larger than the bulk phase coherence length. Furthermore, for a 500\,nm wide nanoribbon, Shubnikov--de Haas (SdH) oscillations are identified at high magnetic fields. The two-dimensional sheet carrier concentration as well as the mobility deduced from the SdH oscillations is superior to the bulk values obtained from Hall measurements. The highly mobile two-dimensional sheet can be coined topological since a Berry phase offset of $\beta\sim \pi$ has been determined.\\

\begin{figure*}[!htb]
    \centering
    \includegraphics[width=\textwidth]{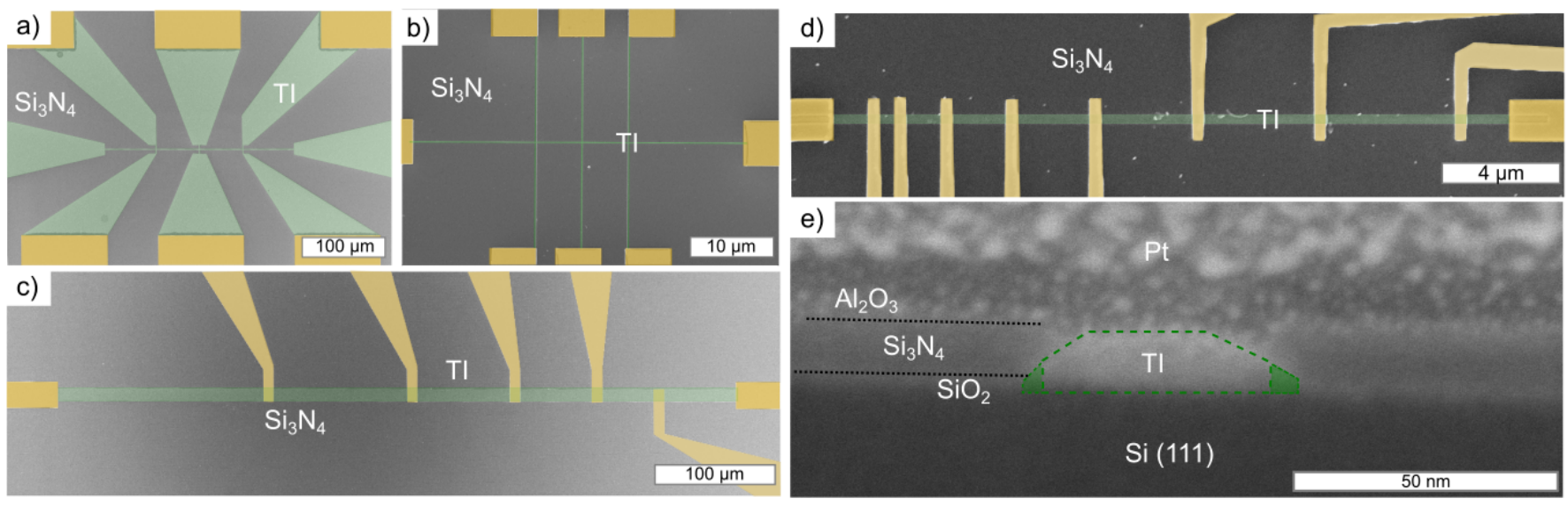}
    	\caption{Selectively grown Bi$_2$Te$_3$ devices. Scanning electron micrographs of a) a 1\,$\mu$m wide micro Hall bar, b) a 50\,nm wide nano Hall bar, c) a 10\,$\mu$m wide TLM microribbon, and d) a 200\,nm wide TLM nanoribbon. The devices are defined in deep-etched trenches within a layer stack of 20\,nm Si$_3$N$_4$ on top of 5\,nm SiO$_2$, i.e. the SAG mask. Planar Ti/Au (20\,nm/50\,nm) contacts have been deposited and are shown in gold/orange. In e) a cross section of a 50\,nm wide nanoribbon is exemplarily shown, which is prepared using focused ion beam milling. The cross section shows the layer stack of the SAG mask, the Al$_2$O$_3$ capping layer, as well as the roughly rectangular nanoribbon cross section. Due to isotropic wet etching of the SiO$_2$ triangular pockets form below the Si$_3$N$_4$ layer. The TI will continue growing in these pockets, which are highlighted in shaded green. The Pt layer is deposited prior to ion beam milling only.}
    	\label{fig1}
\end{figure*}

\section{II. Experimental and Methods}
\subsection{A. Mask Fabrication and Selective Area Growth}
TI nanoribbons and Hall bars have been grown by MBE following a selective-area growth (SAG) approach. Bi$_2$Te$_3$ crystallizes in the tetradymite structure in so-called quintuple layers, where two consecutive quintuple layers are connected by van der Waals forces. The first 5\,nm of a Si(111) 2000\,$\Omega\cdot$cm wafer are thermally converted into SiO$_2$ using a tempress furnace. Subsequently a 20\,nm thick Si$_3$N$_4$ layer is deposited via low-pressure chemical vapor deposition. The wafer is then covered with AR-P 6200 (CSAR) positive electron beam resist and the desired structures are defined. Using reactive ion etching (CHF$_3$/O$_2$ gas mixture) and hydrofluoric acid wet etching, the Si$_3$N$_4$ and the SiO$_2$ layers are etched, respectively. Thus the Si(111) surface is locally revealed, where the TI is to be grown selectively. The structured Si$_3$N$_4$/SiO$_2$ layers therefore form the SAG mask. In  Figs.~\ref{fig1}a) - d) selectively deposited TI devices are shown, i.e. a micro Hall bar, a nano Hall bar, a TLM microribbon, and a TLM nanoribbon, respectively. A scanning electron micrograph of the cross section of a nanoribbon, prepared by focused ion beam milling, is presented in Fig.~\ref{fig1}e). The SiO$_2$ etches isotropically in the hydrofluoric acid used. Therefore, the Si$_3$N$_4$ layer is slighlty underetched. In these pockets the TI will continue to grow, highlighted in shaded green in Fig.~\ref{fig1}.\\

The standard parameters for selective growth of Bi$_2$Te$_3$, given a substrate temperature of T$_{\text{sub}}$=300\,\degree C a Bi-cell temperature of T$_{\text{Bi}}$=470\,\degree C and a Te-cell temperature of T$_{\text{Te}}$=325\,\degree C, result in a growth rate of 7\,nm/h. The TI films are grown in the Te-overpressure regime, since it does not influence the stochiometry but keeps Te from desorbing during growth.\\

\subsection{B. Electrical Measurement Setup}
In order to define ohmic contacts a 50K/50K/950K PMMA three-layer resist stack is used. After development, before deposition of the metal, the sample is immersed for 70\,s in a MF-CD 26 alkaline developer to locally remove the Al$_2$O$_3$ capping layer. Finally, the Ti/Au ohmic contacts are deposited in a vacuum chamber by means of thermal evaporation.\\
\indent Magnetotransport measurements have been performed in a \ch{^{3}He}/\ch{^{4}He} dilution refrigerator with a base temperature of 17\,mK and a \ch{^{4}He} variable temperature insert (VTI) cryostat with a base temperature of 1.5\,K. TLM nanostructures have been characterized in the dilution refrigerator, which can provide magnetic fields of up to 6\,T perpendicular out-of plane and up to 1\,T parallel and perpendicular in-plane. Nano Hall bars have been characterized in the VTI cryostat. The maximum magnetic field provided measures 13\,T. The sample rod for the VTI cryostat is equipped with a rotatable substrate holder, in order to be able to rotate the sample with respect to the static magnetic field. TLM structures have been characterized using a standard two-terminal configuration. For the nano Hall bars two- as well as four-terminal measurements have been performed. For current injection, standard a.c. lock-in techniques have been used.\\

\section{III. Results and Discussion}

\subsection{A. TLM Structures}
\begin{figure*}[!htb]
	\centering
\includegraphics[keepaspectratio=true,width=0.9\textwidth]{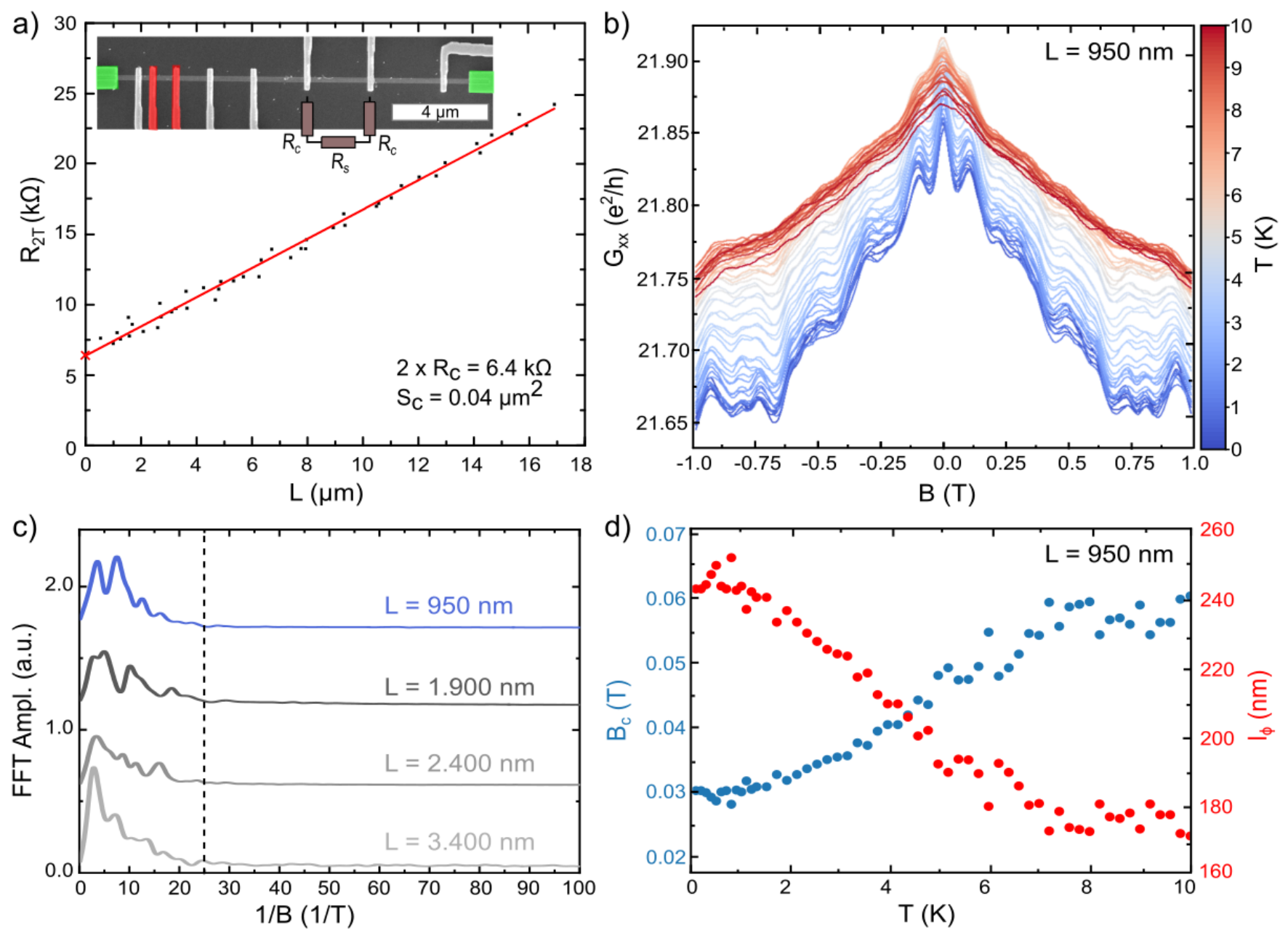}
	\caption{Nanoribbon with applied Ti/Au contacts of different contact separation length $L$. a) TLM measurements performed to determine the contact resistance $R_c$ as well as the sheet resistance $R_S$ and the bulk resistivity $\rho$. The red line is the linearly extrapolated fit to the two-terminal resistance measurements performed in between every alteration of contacts. A schematic of the measurement scheme as well as an SEM image of the 200\,nm wide TLM nanoribbon are shown in the inset. b) Temperature-dependent magnetoconductance measurements $G_{xx}(B)$ performed on a 950\,nm long nanoribbon segment highlighted in a). Measurements have been performed in a four-terminal setup (green: current bias, red: potential difference). c) FFT performed on the magnetoconductance $G_{xx}$ on different segments of the nanoribbon. The blue curve highlights the FFT performed on the data at 17\,mK base temperature shown in b). d) Results of the correlation field $B_c$ analysis, as well performed on the data at 17\,mK base temperature shown in b).}
	\label{fig2}
\end{figure*}

TLM \cite{Berger1969,Marlow1982} measurements were performed on a 200\,nm wide nanoribbon, to determine the contact resistance $R_c$ of the Ti/Au-Bi$_2$Te$_3$ interface, the sheet resistance $R_s$ as well as the bulk resistivity $\rho$ of the selectively grown Bi$_2$Te$_3$ nanoribbon. For comparison, TLM measurements were also performed on a Bi$_2$Te$_3$ microribbon with a width of 10\,$\upmu$m. A scanning electron micrograph of the nano TLM device is shown in the top left corner of Fig.~\ref{fig2} a). Two-terminal measurements of the total resistance $R_{\text{2T}}$ are performed in between every possible combination of contacts. The total resistance in between two adjacent contacts is given by
\begin{equation}\label{TLM}
    R_{\text{2T}} = \frac{L}{w} \cdot R_S + 2 \cdot R_c \; ,
\end{equation}
where $w$ is the width of the nanoribbon. $R_S$ is assumed to be constant throughout the whole ribbon. Furthermore, $R_{2T}$ is assumed to be linearly dependent on $L$, as the nanoribbon has a constant width. For measured values of $R_{\text{2T}}$, the total resistance of the d.c. lines as well RC-filter elements in series to the TLM nanoribbon have been subtracted. The results are displayed as a function of the contact separation distance $L$ in Fig.~\ref{fig2} a). From the slope of the $R_{2T}(L)$ measurements the sheet resistance of the Bi$_2$Te$_3$ nanoribbons has been deduced.  With a slope of $R_S/w=1.04\,\Omega$/nm the sheet resistance is obtained as $R_S=208\,\Omega$ and the bulk resistivity as $\rho=R_S\cdot t =3.5\times10^{-4}\,\Omega\cdot$cm. In comparison, for the TI microribbon a slope of $R_S/w=31.25\,\Omega/\upmu$m results in $R_S=312.5\,\Omega$ and $\rho=R_S\cdot t =5.3\times10^{-4}\,\Omega\cdot$cm. Thus, the bulk resistivity of the nanometer and the micrometer wide ribbons are comparable.\\

Following Eq.~\ref{TLM} the contact resistance of both contacts $2\cdot R_c$ result in an offset to the linear relation of the total resistance and the contact separation. The contacts resistance can therefore be extracted by extrapolating $R_{\text{2T}}(L)$ and determining the y-intercept (see red line Fig.~\ref{fig2} a)). The single contact resistance $R_c$ for the 200\,nm wide TLM ribbon is determined to be $R_{c,\mathrm{nano}}=3.2\,$k$\Omega$, whereas for the 10\,$\upmu$m wide ribbon we determined a value of $R_{c,\mathrm{micro}}=200$\,$\Omega$, which gives a ratio of $R_{c,\mathrm{nano}}/R_{c,\mathrm{micro}}=16$. The contact area $S_{c,\mathrm{micro}}$ of the micro TLM device measures 100\,$\upmu$m$^2$. For comparison, the contact area $S_{c,\mathrm{nano}}$ of the nano TLM ribbon measures 0.04\,$\mu$m$^2$ resulting in $S_{c,\mathrm{micro}}/S_{c,\mathrm{nano}}=2.500$. Thus, despite the much smaller contact area, the interface resistance stays sufficiently small in order to study magnetotransport phenomena using as prepared Ti/Au contacts.\\

The TLM nanoribbon device has been used to study quantum modulations to the macroscopic magnetoconductance as a function of the device length $L$.  When a magnetic field is applied perpendicular to the nanoribbon, a sharp weak antilocalization (WAL) feature in between $-0.1\,$T $ \leq B \leq +0.1\,$T as well as additional universal conductance fluctuations (UCFs) for $B > \pm 0.1\,$T are observed for each segment of the nanoribbon. The WAL effect as well as UCFs both originate from interference of partial electron waves. This occurs within defect-based electron paths in real space that form closed loops. An applied magnetic flux can thereby shift the phase of the partial waves and affect the interference pattern. Therefore, WAL and UCF features can both be used to determine the phase-coherence length $l_{\phi}$ as quantum coherence is the major prerequisite for the observation of both effects.\\
\indent In Fig.~\ref{fig2} b) temperature dependent magnetoconductance measurements, performed in a four-terminal configuration, are shown. The two outermost contacts (displayed in green in the inset of Fig.~\ref{fig2} a)) have been used to apply a d.c. current bias of 10\,nA. Two neighbouring contacts (displayed in red in the inset of Fig.~\ref{fig2} a)) have been used to measure the potential difference in this nanoribbon segment of $L=950\,$nm. For temperatures near 10\,K the amplitude of the WAL and UCF features are strongly suppressed, but possible to identify. An FFT performed on the dataset obtained at 17\,mK (as highlighted in blue in Fig.~\ref{fig2} c)) shows several prominent frequencies. In the same graph FFTs performed on different segments of the nanoribbon are shown (highlighted in different shades of grey). See supplemental material at \cite{Supplementary} for temperature dependent magnetoconductance measurements on these other segments of the nanoribbon. As is typical for UCFs, each and every segment of the nanoribbon shows different prominent frequencies in the FFT. However, also for each segment above 25\,T$^{-1}$, no more prominent frequencies have been observed. Following $A=\Phi_0 /B$, this threshold value corresponds to a maximum area of $A_{th}=5.2\times 10^{-14}\,$m$^2$. Loops of phase-coherent transport of this size maximally span perpendicular to the applied magnetic field. The maximum area determined from the FFT, devided by the nanoribbon width $w$, gives an estimate of the phase-coherence length $l_{\phi}$ along the ribbon. This only holds under the assumption, that the nanoribbon width is shorter than the phase-coherence length $l_{\phi}$. In fact $l_{\phi} \sim l=A_{th}/w=260\,$nm exceeds the junction width.\\
\indent In order to determine $l_{\phi}$ more explicitly a temperature dependent correlation field analysis of the form
\begin{equation}
F(\Delta B) = \langle \delta G (B+\Delta B)\delta G(B)) \rangle
\end{equation}
on the same data is performed \cite{Beenakker1991,Lee1987,Beenakker1988}. In the dirty limit, the relation between $B_c$ and $l_{\phi}$ is expressed by $B_c(l_{\phi})=\gamma \frac{e}{h} \frac{1}{wl_{\phi}}$, where the prefactor $\gamma$ is chosen to be 0.42  \cite{Beenakker1991}. Results for the correlation fields $B_c$ and the corresponding phase-coherence lengths $l_{\phi}$ are shown in Fig.~\ref{fig2} d). Interestingly, $l_{\phi}$ at base temperature coincides with the phase-coherence length estimated from the FFT performed. Furthermore, it can be seen, that it does not increase anymore below $T \sim 2\,$K. When considering only electron-electron interaction, the phase coherence length should increase even further for lower temperatures \cite{Wang2011}. Scattering at the physical boundaries of the nanoribbon, might be the reason that no increase of $\l_{\phi}$ is observed for lower temperatures. Therefore, width dependent nano Hall bar measurements are performed in the VTI, equipped with a 13\,T superconducting magnet. These measurements are discussed in the next section.

\subsection{B. Nano Hall Bars}

\begin{figure*}[!htb]
	\centering
\includegraphics[width=\textwidth]{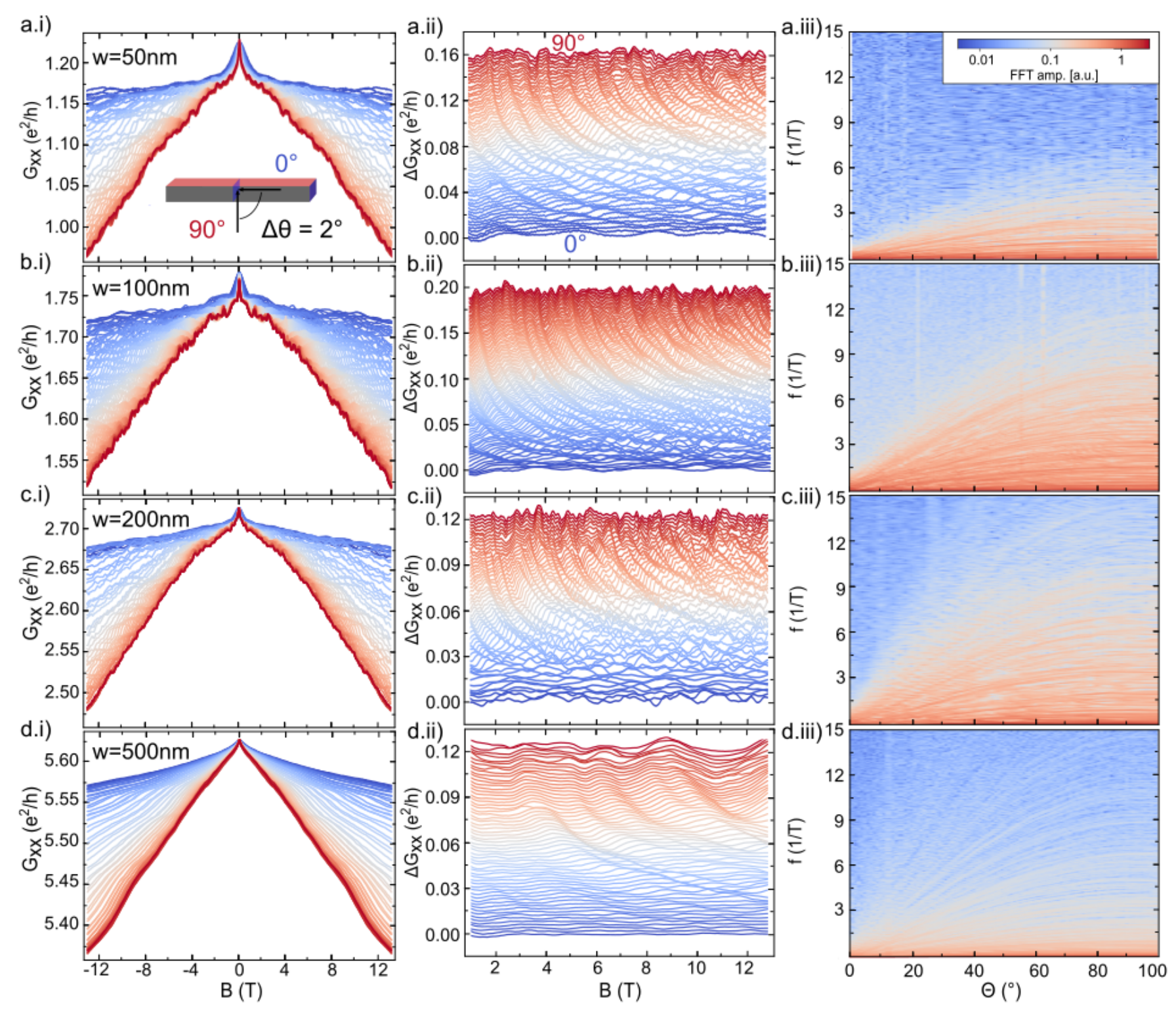}
	\caption{Angle dependent measurements performed on the selectively grown Bi$_2$Te$_3$ nano Hall bars of different width $w$. a.i)-d.i) Angle dependent magnetoconductance measurements. The red curves show the magnetoconductance behavior at an effectively perpendicular out-of-plane magnetic field and the blue curves at an effectively parallel in-plane magnetic field as shown in the schematic in a.i). The gradient in color in between the red and blue curves show intermediate angles. a.ii-d.ii) show the magnetoconductance values after subtraction of an averaging background. For clarity, only positive magnetic field values, excluding the WAL feature below 1\,T, are plotted. a.iii)-d.iii) FFT amplitude as a function of tilt angle and frequency. The FFT is performed on the whole data range of $-13\,\mathrm{T} \leq B \leq +13\,\mathrm{T}$.}
	\label{fig3}
\end{figure*}

Four nanoscale Hall bars have been characterised in the VTI at 1.5\,K. The Hall bars consist of one main nanoribbon with a width of 50\,nm (exemplarily shown in Fig.~\ref{fig1} b)), 100\,nm, 200\,nm or 500\,nm. Along this nanoribbon a d.c. current bias is applied. The main ribbon extends into six nanoribbon contacts in order to measure the potential difference along or across the main ribbon. Bulk sheet carrier density values are determined from Hall measurements to be in the range of $n_{2D}=(6.0-9.5) \times 10^{13}\,$cm$^{-2}$. Similarly, bulk mobility values are determined to be in the range of $\upmu = (180-210)\,$cm$^2$/Vs. Bulk sheet carrier concentrations as well as bulk mobility values determined on 1\,$\upmu$m and 10\,$\upmu$m wide selectively-grown Hall bars lie within the same range as has been determined for the nano Hall bars. These values on our selectively grown, Bi$_2$Te$_3$ 3D TI nanoribbons deem the bulk to have a rather metallic character. See supplemental material at \cite{Supplementary} for a detailed study of the Hall measurements. \\
\indent Four-terminal magnetoconductance measurements on the nano Hall bars however, show some distinct features, when compared to the micrometer wide Hall bars. For the longitudinal conductance $G_{xx}$ of the nano Hall bars, a similar behavior as determined within the TLM nanostructures is found. Fig.~\ref{fig3} a.i)-d.i) (red curves) shows that an applied perpendicular magnetic field results in sharp WAL features with symmetric UCF modulations superimposed onto a mesoscopic background. With regards to FFTs performed, the 13\,T range of the superconducting magnet results in a much better resolution, when compared to previous measurements on the TLM nanoribbons with a range of 1\,T ($\Delta \frac{1}{B}=1/B_{max}=\frac{1}{14}\,$T$^{-1}=0.071\,$T$^{-1}$).\\

The Hall bar can be tilted from an effectively perpendicular magnetic field ($\Theta=90^{\circ}$, Figs.~\ref{fig3} a.i)-d.i), red curves) to a field parallel to the current flow ($\Theta=0^{\circ}$, Figs.~\ref{fig3} a.i)-d.i), blue curves) using a rotatable sample holder. The magnetoconductance measurements performed with a stepping of $\Delta \Theta = 2^{\circ}$ show a systematic change in the modulation pattern. This systematic change becomes more evident, when having a look at the background-subtracted magnetoconductance data $\Delta G_{xx}$ within the range of $1\,$T $ \leq B \leq 13\,$T, as shown in Figs.~\ref{fig3} a.ii)-d.ii). The position of most of the quantum modulation peaks were found to follow a sinusoidal behavior as a function of the tilt angle $\Theta$. The following discussion is based on reference \cite{Koelzer2019}. Maps of the FFT amplitude as a function of frequency ($1/B$) and tilt angle ($\Theta$) are shown in Figs.~\ref{fig3} a.iii)-d.iii). The above mentioned sinusoidal dependency of the oscillation frequency of a single defect based coherent interference loop becomes even more evident in the FFT maps. In their origin these loops however are just projections of any complex 3D electronic path onto the relevant 2D plane, perpendicular to the applied magnetic field \cite{Beenakker1991}. The overall flux within the loop is therefore expressed by
\begin{equation}
    \Phi = B_x S_{yz} + B_y S_{xz} + B_z S_{xy},
\end{equation}
where $B_x$, $B_y$ and $B_z$ are the magnetic field components of any arbitrary 3D magnetic field vector and $S_{yz}$, $S_{xz}$ and $S_{xy}$ are the 2D projections of any arbitrary 3D coherent electronic interference path. For interference paths, that would solely expand on a 2D plane, the maximum observed frequency $1/B$ lies at exactly $\Theta=90^{\circ}$. Furthermore, in this scenario exactly at a 90\,$^{\circ}$ offset from that maximum frequency, the effective flux through the 2D interference loops should be 0. In the measured data, the maximum at $\Theta = 90\,^{\circ}$ is observed for every Hall bar. Many traceable features are observed, indicating a few prominent interference loop sizes. However, at $\Theta = 0\,^{\circ}$ quantum modulations with finite frequency are still observed in the FFT. The reason therefore is that at low angles Aharonov--Bohm (AB) type oscillations are expected to originate from the topologically protected surface states encircling the nanoribbon cross section \cite{Peng2010,Zhang2010,Xiu2011,Cho2015,Jauregui2015,Arango16}. At low angles ($\Theta \leq 10^{\circ}$) both AB type oscillations as well as UCFs are visible in the magnetoconductance. Similar findings have previously been reported \cite{Arango16,Haas2016} and can as well be deduced from angle dependent measurements on the TLM nanoribbon structure. See supplemental material at \cite{Supplementary} for a discussion of the angle dependent measurements on the TLM nanoribbon. The AB type modulations to the magnetoconductance are analysed in the following section.\\

\subsection{C. Aharonov--Bohm Oscillations}

\begin{figure*}[!htb]
	\centering
\includegraphics[width=\textwidth]{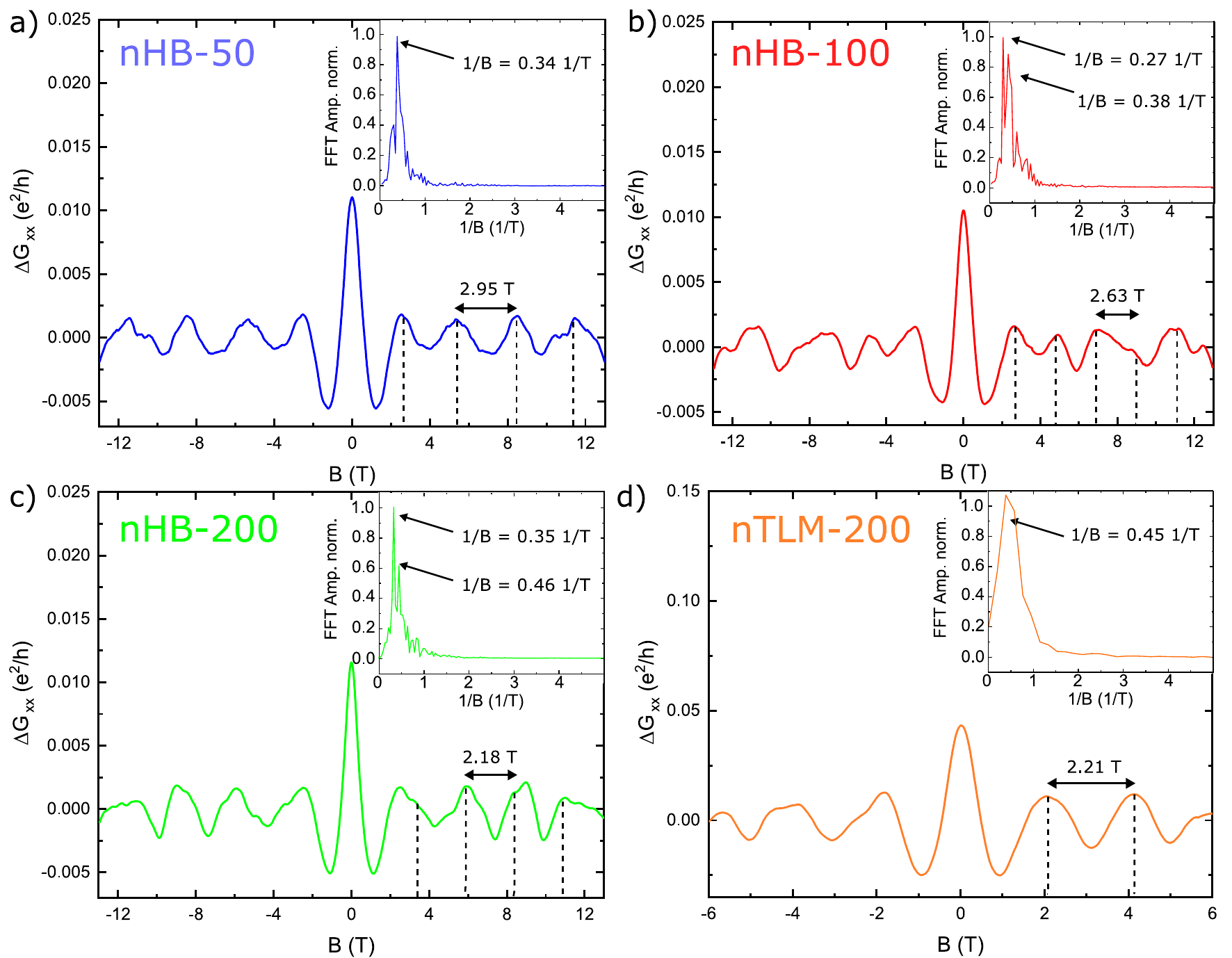}
	\caption{Magnetoconductance measurements in a magnetic field applied along the nanoribbon axis. The a) 50\,nm (nHB-50), b) 100\,nm (nHB-100) and c) 200\,nm wide nano Hall bar, as well as one of the d) 200\,nm (nTLM) wide nano TLM ribbon segments, are compared. A smooth background has been subtracted and the change in conductance $\Delta G_{xx} (B)$ is shown. For $B > \pm 1\,$T several pronounced AB oscillation periods can be identified. An FFT is performed and shown in the inset of each figure. In the FFT peaks with highest amplitude are identified and frequencies in $1/B$ mentioned.}
	\label{fig5}
\end{figure*}

AB type modulation features at small angles $\Theta \leq 10^{\circ}$ are observed in the nano TLM and the nano Hall bar structures. The expected frequencies ($f=1/B$) for interference after traversing the whole nanoribbon cross section $S=S_{xy,max}$ are determined using $f = eA/h$. An overview of the expected AB oscillation frequencies is given in Tab.~\ref{Tab1} together with the nanoribbon dimensions. Magnetoconductance measurements at fields applied along the nanoribbon axis have been performed on the 50\,nm (nHB1), 100\,nm (nHB2) and 200\,nm (nHB3) wide Hall bars as well as on the 200\,nm wide TLM nanoribbon segments (nTLM).\\
\indent In order to isolate the AB oscillations ($\Delta G_{xx} (B_z)$) from the macroscopic magnetoconductance a smooth background has been subtracted by applying a first order Savitzky--Golay filter. The results are shown in Figs.~\ref{fig5} a)-d). The prominent features in between $-1\,$T $\leq B \leq +1\,$T are explained by the WAL effect. Outside this range the different curves all show a distinct oscillatory behavior, each with prominent oscillation periods $\Delta B$. The period of these oscillations are highlighted within each graph. FFT results on each data set are shown in the respective inset. The amplitude of the frequency count (normalized to the greatest amplitude determined) is displayed  as a function of the frequency $1/B$. The most prominent frequencies are marked in every graph. It has to be noted, that in the dilution refrigerator, where the 200\,nm wide TLM nanoribbon segment (nTLM-200) has been measured, only a maximum magnetic field of 6\,T could have been applied. This maximum field lowers the possible resolution of the FFT performed to $\Delta (1/B) = 1/6\,$T$^{-1}=0.167\,$T$^{-1}$.\\

\begin{table}[!htb]
\begin{tabular}{| c | c | c | c | c | c |}
\hline
device & $w$ [nm] & $t$ [nm] & $S$ [nm$^2$] & $S\frac{e}{h}$[1/T] & $f_{FFT}$[1/T] \\
\hline
\hline
  nHB-50 & 50 & 20 & 1000 & 0.24 & 0.34 \\
  \hline
  nHB-100 & 100 & 18 & 1800 & 0.44 & 0.38 \\
  \hline
  nHB-200 & 200 & 14 & 2800 & 0.68 & 0.46 \\
  \hline
  nHB-500 & 500 & 10 & 5000 & 1.21 & - \\
  \hline
  nTLM-200 & 200 & 14 & 2800 & 0.44 & 0.45 \\
 \hline
\end{tabular}
\caption{Overview of the geometry of the five different nanoribbon devices investigated. The cross sectional areas $S$ are determined and the expected AB frequencies are calculated. For comparison, the frequencies $f$ determined from the FFTs performed are given as well.}
\label{Tab1}
\end{table}

The 50\,nm Hall bar shows a frequency count with greatest amplitude at $f=1/B=0.34\,$T$^{-1}$ ($\Delta B=2.94\,$T). The corresponding area measures  $e/hB=1157\,$nm$^2$. This area determined using FFT is bigger than the geometrically identified area. For the 100\,nm Hall bar two prominent features are observed at $f=0.27\,$T$^{-1}$ ($\Delta B=3.70\,$T, $e/hB=1116\,$nm$^2$) and at $f=0.38\,$T$^{-1}$ ($\Delta B=2.63\,$T, $e/hB=1570\,$nm$^2$). Prominent frequencies observed for the 200\,nm Hall bar are $f=0.35\,$T$^{-1}$ ($\Delta B=2.86\,$T,$e/hB=1447\,$nm$^2$) and at $f=0.46\,$T$^{-1}$ ($\Delta B=2.17,$T, $e/hB=1902\,$nm$^2$).  In case of the 100\,nm as well as 200\,nm wide Hall bar the areas determined from the FFT are both smaller than the geometrically identified areas. However, when the 200\,nm wide Hall bar is compared to the 200\,nm wide TLM nanoribbon segment  ($f=0.45$T$^{-1}, \Delta B=2.22\,$T$, e/hB=1902\,$nm$^2$), both determined frequencies from correpsonding FFTs match, as highlighted in Tab.~\ref{Tab1}.\\
\indent It is possible that the geometrically identified areas for the 100\,nm and 200\,nm wide ribbons are overestimated. It has been observed that for wider ribbon cross sections ($w > 100\,$nm) the average film thickness is lower than for less wide junctions. The reason is the SAG process used. Given a certain mobility of adatoms during MBE growth, an adatom diffusion length can be defined. Inside the nanotrenches not only adatoms directly impinging onto the Si(111) surface but also collected from the nearby amorphous nitride surfaces will be used for TI growth. The adatom diffusion length defines an area around the edges of the nanotrench, from which these additional adatoms will be collected. For thinner nanotrenches the ratio of this area to the surface area of the exposed Si(111) is bigger than for wider nanotrenches. Therefore, the material within thin nanotrenches has the tendency to grow thicker. Furthermore, due to the isotropic etching of SiO$_2$ during SAG mask fabrication, the Bi$_2$Te$_3$ layers extend underneath the silicon nitride, as explained in the method section. These pockets might be considered to determine the effective area of the 50\,nm wide Hall bar but might not be considered for wider nanoribbons.\\
\indent The prominent frequencies determined from the FFT analysis on the magnetoconductance data highlight the existence of AB oscillation patterns originating from only a few fixed loop sizes in the nanoribbon cross sections. The largest amplitude is observed for frequencies that correspond to loop sizes matching the cross sectional area of the nanoribbons. Periodic oscillations as can be seen in the background subtracted magnetoconductance data, correspond to these frequencies. The circumference of the 200\,nm wide Hall bar however already exceeds the bulk phase coherence length $l_{\phi}$ of the nanoribbon ($U_{100\,\text{nm}}=240\,$nm$<l_{\phi}<U_{200\,\text{nm}}=440\,$nm) previously determined. A possible explanation might therefore be an additional, high coherent trajectory spanning across the nanoribbons surfaces, that is partially decoupled from the bulk. However, it is not unambiguously distinguishable, whether these AB oscillations originate from topological surface states or classical two-dimensional surface states. A classical two-dimensional, highly conductive sheet can result from band bending effects at the different surfaces \cite{Frantzeskakis17}. A distinction might be drawn by extracting the Berry phase from Shubnikov--de Haas oscillations that would indicate the surface states to be of topological nature \cite{Mikitik1999}. Consequently, Shubnikov--de Haas (SdH) oscillations observed in a magnetic field applied perpendicular out-of plane are analysed and discussed in the next section.\\

\subsection{D. Shubnikov--de Haas Oscillations}
The angular dependent magnetoconductance of the 500\,nm wide Hall bar shows some distinctive features, when compared to the 50\,nm, 100\,nm and 200\,nm wide Hall bars. In the in-plane parallel field orientation, the highest observed frequency corresponds to an area much smaller than the cross section of the nanoribbon. The circumference of the nanoribbon ($U_{500\,\text{nm}}=1020\,$nm) seems to be larger than the phase coherence length. In the out-of plane perpendicular field orientation, high frequency UCF spectra are observable in the FFT analysis (Fig.~\ref{fig3} d.iii)) but with reduced relative amplitude. The most apparent difference however is the low-frequency SdH oscillations that do show up in the perpendicular field orientation at high magnetic fields $B\geq 4\,$T. These can be observed in the $G_{xx} (B,\Theta)$ data of the 500\,nm wide Hall bar only. In the $\Delta G_{xx} (B,\Theta)$ dataset (Fig.~\ref{fig3} d.ii)) it is evident, that this feature shifts with the sine of the angle $\Theta$. The SdH oscillations therefore originate from a 2D sheet parallel to the sample surface.

\begin{figure*}[!htb]
	\centering
\includegraphics[width=\textwidth]{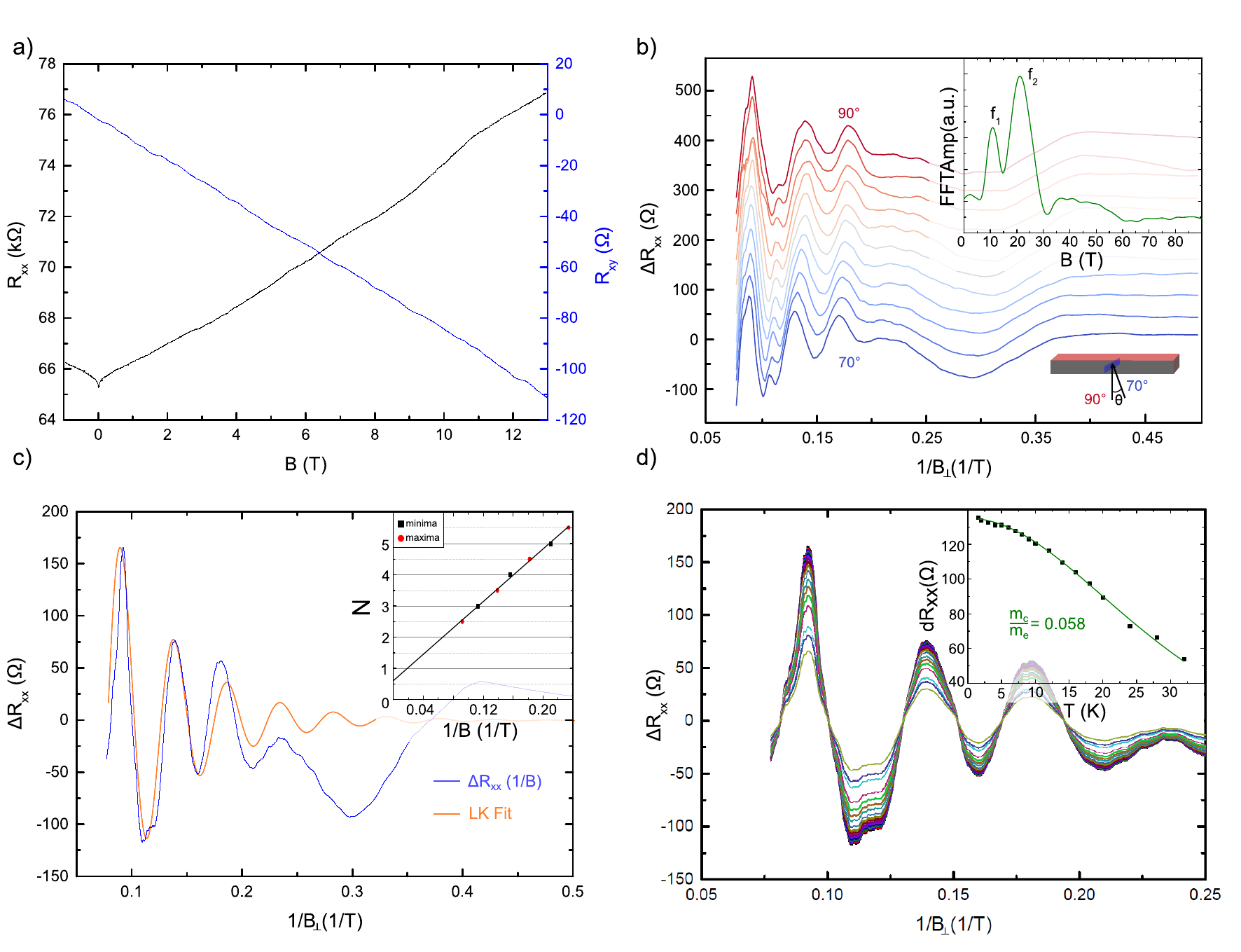}
	\caption{Analysis of low frequent SdH-oscillations observed for the 500\,nm wide Hall bar. a) Longitudinal $R_{xx}$ (black curve) as well as Hall resistance $R_{xy}$ (blue curve) measured at $\Theta=90^{\circ}$. The longitudinal resistance shows some long frequent oscillations at higher fields $B \leq 5$\,T. In b) these oscillations are plotted after background subtraction as a function of the inverse perpendicular magnetic field $\Delta R_{xx} (B_{\perp}$ for angles in between $70^{\circ} \leq \Theta \leq 90^{\circ}$). The FFT shown in the inset is performed on the $\Theta = 90^{\circ}$ sweep and shows two prominent frequencies $f_1=11.3$\,T and $f_2 = 21.7$\,T. c) The general LK expression (orange curve) is fitted to the SdH-oscillations (blue curve) to extract mobility $\mu=1997.3\,\frac{cm^2}{V\cdot s}$ and phase-offset $\beta_{LK}=0.508$. In the inset of c) the maxima and minima in the SdH-oscillations are plotted against the inverse magnetic field as half integer and full integer Landau levels $N$, respectively. The linear fit performed is used to extract the y-intercept (phase-offset $\beta_{LL}=0.612$). d) Temperature dependent SdH-oscillations pattern. The oscillation amplitude at $1/B=0.0921\,1/T$ is plotted in the inset as a function of temperature. A fit based on the thermodynamic part of the Lifshitz-Kosevich expression is used to extract the cyclotron frequency $\omega_c=3.19 \times 10^{13}$\,1/s.}
	\label{fig4}
\end{figure*}

In Fig.~\ref{fig4} a) the longitudinal resistance $R_{xx}(B)$ for an applied magnetic field perpendicular ($\Theta=90^{\circ}$) to the 500\,nm wide nanoribbon is shown. $R_{xx}(B)$ is displayed together with the Hall voltage $R_{xy}(B)$. Next to a sharp WAL in $R_{xx}(B)$, low frequent SdH oscillations can be seen at higher magnetic fields. A smooth background using a Savitzky--Golay filter is subtracted from the longitudinal magnetoresistance $R_{xx}(B)$ curve. The result is shown as a function of the inverse magnetic field $\Delta R_{xx}(1/B)$ in Fig.~\ref{fig4} b) (dark red curve) and shows periodic SdH magnetoresistance modulations. For comparison, the $\Delta R_{xx}(1/B_{\perp})$ values are displayed for various tilt angles in between $70^{\circ}\leq \Theta \leq 90^{\circ}$. For the angle dependency of the SdH-oscillations only the perpendicular magnetic field component ($B_{\perp} = B \cdot \sin{\Theta})$ has been considered. The observed oscillation period does not shift upon changing the angle $\Theta$. The FFT performed on the curve at $\Theta =90^{\circ}$ (Fig.~\ref{fig4} b), insert) shows two distinct frequencies $f_1=11.3$\,T and $f_2=21.7$\,T. The second frequency matches the period observed in the $\Delta R (1/B_{\perp})$ magnetoresistance curve. Oscillation frequency $f_1$ relates to the same oscillations with doubled period. Using the Onsager relation $f= e/2n_\mathrm{SdH}\pi\hbar$  \cite{Weyrich2017}, with $f=f_2$ being the frequency of the SdH-oscillations, we can extract a 2D sheet carrier concentration of $n_{\mathrm{SdH}}=5.31 \times 10^{11} $cm$^{-2}$, for $f=f_2=21.7\,$T. This value can be directly compared to the Hall measurements performed on this 500\,nm wide Hall bar $n_{2D,\mathrm{Hall}}=7.55 \times 10^{13}\mathrm{cm}^{-2}$. It becomes apparent that the sheet carrier concentration extracted from the SdH-oscillations is about two orders of magnitude smaller than that from the Hall measurements. It is therefore apparent, that there is a transport channel, which is at least partially decoupled from the bulk of the material.

In order to evaluate if that additional 2D sheet originates from possible topological surface(s), the Lifshitz--Kosevich (LK) expression  \cite{Akiyama_2018,Okazaki2018,Xiong2012} of the form,
\begin{equation}
    \Delta R_{xx} = a \cdot e^{-\pi / \mu \cdot B} \cdot \cos{\left( 2\pi \cdot \frac{f}{B}+ \pi +\beta_{LK} \right)}
\end{equation}
is used to fit $\Delta R_{xx}(1/B_{\perp})$ (see Fig.~\ref{fig4} c), pink curve). In the equation above $a$ is the oscillation amplitude, $\mu$ is the mobility, $f=f_2$ is the oscillation period and $\beta_{\mathrm{LK}}$ is the phase-offset. In the fit performed $\mu$ and $\beta$ are used as fit parameters and results in a mobility value of $\mu_{\mathrm{SdH}}=1997.3$\,$\frac{\mathrm{cm}^2}{\mathrm{V}\cdot \mathrm{s}}$ and a phase offset of $\beta_{LK} = 0.507$. The mobility $\mu_{SdH}$ is thereby also found to be about a factor of 200 larger than the bulk mobility determined from the Hall measurements $\mu_{bulk}=104$\,$\frac{\mathrm{cm}^2}{\mathrm{V}\cdot \mathrm{s}}$.\\

In order to identify the origin of the two-dimensional transport channel with higher mobility $\mu$ and lower sheet carrier concentration $n$ the physically relevant phase-offset $\beta_\mathrm{LK}$ extracted from the LK-fit is considered. For a classical two-dimensional electron gas in an accumulation space charge layer the expected phase-offset would be zero. A deviation from that value has been reported to be the case in linear Dirac systems due to the Berry phase \cite{Mikitik1999}. The spin-momentum locking present in the TIs leads to a Berry phase offset of $\pi$ for a charge traversing one full circle in $k$-space. A phase-offset of $\pi$ has already been reported for other Dirac systems as well  \cite{Akiyama_2018,Xiong2012}. To further evaluate the phase-offset of the Landau level spacing, the Landau level index $N$ is plotted against the inverse perpendicular magnetic field $1/B_{\perp}$ in the inset of Fig.~\ref{fig4} c). All together seven extrema can be identified that match the oscillatory pattern of frequency $f_2=21.7\,$T. The minima are addressed as full integer Landau level steps, while the maxima are addressed as half integer Landau level steps.The phase-offset $\beta_{LL}$ can be identified by interpolating the Landau level fan diagram in order to determine the y-intercept. The phase-offset extracted from the interpolated line is $\beta_{LL}=0.612$, which is in good agreement with the phase-offset extracted from the LK fit ($\beta_{LK} = 0.507$). Thus, it is plausible that the two-dimensional transport channel indeed originates from topological surface states.

By assuming a linear dispersion for the topological surface states identified as $E=v_Fk\hbar$ the Fermi energy can be deduced from the oscillatory frequency of the SdH-oscillations following the relation
\begin{equation}\label{Fermi_Energy}
    E_F = \sqrt{f_2 \cdot v_F^2 \cdot 2\hbar} \; ,
\end{equation}
where $f_2=21.7\,$T and $v_F$ is the Fermi velocity. In order to determine the Fermi velocity $v_F$, the Dirac dispersion relation $\delta E / \delta k = v_F \hbar$ as well as the general expression for the cyclotron mass $m_c=k_F \hbar^2 (\delta E / \delta k)^{-1}$ was used. The Fermi wave vector is given by $k_F=\sqrt{4\pi n_{SdH}}=2.58\times10^8\,$m$^{-1}$, assuming a 2D space. The cyclotron mass was obtained from a fit based on the temperature-dependent part of the LK expression \cite{Weyrich2017} to the temperature-dependent oscillation amplitude of the SdH-oscillations. In Fig.~\ref{fig4} d) $\Delta R_{xx}(1/B_{\perp})$ for various temperatures $1.5\,\mathrm{K} \leq T \leq 30\, \mathrm{K}$ is shown. The oscillation amplitude of the first maxima at $1/B_{\perp}=0.0921\,1/T$ is plotted as a function of temperature in Fig.~\ref{fig4}d) (insert). The values for the fit obtained result in $\omega_c=3.19 \times 10^{13}$\,1/s. The cyclotron mass is then given by $m_c=e\cdot B / \omega_c = 5.451\times10^{-32}$\,kg, at $B=1/0.0921\,T=10.86\,T$, which is the position of the first SdH maximum. With these values the Fermi velocity can be determined to be $v_F=\hbar k_F / m_c = 5.08 \times 10^5$\,m/s. The position of the Fermi energy, with reference to the Dirac point is evaluated using Eq.~(\ref{Fermi_Energy}) and results in $E_F=152$\,meV. Comparing this to previously reported angle resolved photoemission spectrum measurements performed on bulk samples of Bi$_2$Te$_3$ \cite{Eschbach2016} this value would allocate the Fermi energy to lie slightly below the bulk conduction band edge.\\

\section{IV. Conclusion and Outlook}
We investigated selectively grown Bi$_2$Te$_3$ nanoribbons and nano Hall bars of varying width in low temperature electrical transport studies. Different quantum transport phenomena have been identified in magnetoconductance measurements. The focus has been set to understand, which of these quantum modulation phenomena can be attributed to the bulk of these diffusive nanoribbons and which to the topologically protected surfaces.\\
TLM measurements have been performed on nano- as well as microribbons with Ti/Au contacts in order to evaluate the contact resistance for different contact areas. It has been concluded that even though the contact area for the nanoribbon devices is just a small fraction of the contact area provided for the microribbons ($S_{c,\mathrm{micro}}/S_{c,\mathrm{nano}}=2.500$), the contact resistance is not increased significantly. The sheet resistance as well as the specific bulk resistivity for different ribbon-dimensions has been determined. Including Hall bar measurements performed on both nano- as well as microribbons it can be concluded that the TI ribbon properties, with respect to bulk resistivity, bulk sheet carrier concentration, as well as bulk mobility shows geometric dependency on geometry.\\
\indent By evaluating WAL features and UCF spectra the bulk phase coherence length along the nanoribbon was determined to be $l_{\phi}=240\,$nm. For an out-of plane perpendicular magnetic field this results in traceable defect based coherent electron interference loops, which are limited in one direction by the nanoribbon width. By performing angle dependent magnetoconductance measurements, these interference loops have been determined to reside on two-dimensional planes parallel to the sample surface, mainly originating from the bulk. For a magnetic field parallel to the nanoribbon axis, AB type oscillations have been determined that can be attributed to an additional, highly coherent transport sheet on the circumference of the nanoribbons, that is partially decoupled from the bulk. The angle dependent magnetoconductance measurements furthermore show how the UCFs and the AB type oscillations merge at low angles. The UCF spectra dominate for $\Theta > 10^{\circ}$, while the AB type oscillations dominate for $\Theta \leq 10^{\circ}$. For the 500\,nm wide nanoribbon Shubnikov--de Haas oscillations have been observed. The deduced sheet carrier density is lower and the mobility higher, when compared to the bulk values determined. A Lifshitz--Kosevich fit to the SdH oscillation period in the inverse perpendicular magnetic field has been performed. The phase-offset determined from this fit coincides with the phase-offset determined by analyzing individual Landau levels and measures $\beta \sim \pi$. This phase offset of nearly $\pi$ is a clear indication of the Berry phase, which is only present in topological surface states.\\
\indent The analysis performed shows that the selectively grown Bi$_2$Te$_3$ nanoribbons suffer from a high bulk carrier density and a low bulk mobility. However, convincing signatures of surface state transport, superimposed onto bulk effects, have been found. The characterization tools presented can be used to further study the behavior of selectively grown 3D TI nanoribbons of more bulk insulating materials. The bulk behavior can further be suppressed in future devices by including electrostatic gating. The scalability of the SAG process used in combination with proper bulk insulating nanoribbons will allow for large scale investigation of two-dimensional TI nanoribbon architectures \cite{Schueffelgen2019}. These can ultimatively be included into topological insulator-superconductor hybrid devices, for the realization of quantum computation schemes based on Majorana fermions.

\section{Author contribution statement}
D.R. initiated the project and performed measurements on the nano-TLM and nano Hall bar devices. N.O. performed measurements on micrometer sized TLM and Hall bar devices. D.R. and A.R.J. have been fabricating the devices in clean room conditions. A.R.J. and G.M. have deposited the selectively grown nanoribbon structures. M.M. and N.O. have been designing the nano- and micro-TLM devices. D.R., J.K. and E.Z. have maintained the operation of the cryogenic setups, the electrical measurement devices and the python based measurement scripts. D.R. and S.B. have performed the focused ion beam cut as well as the micrographical analysis of the nanoribbon cross sections. The project has been supervised and intensively discussed with D.G., H.L. and Th.S.

\section{Acknowledgements}
The authors would like to thank Gunjan P. Nagda for a critical review of contents and proof reading of the manuscript. We would like to thank Herbert Kertz for the technical supervision of the experimental setups. This work was financially supported by the Virtual Institute for Topological Insulators (VITI), which is funded by the Helmholtz Association.

\bibliography{QT_in_TI_Nanoribbons}

\end{document}

\pagestyle{empty}

\begin{center}

\begin{Large}
\textbf{Quantum transport in topological surface states of Bi$_2$Te$_3$ nanoscale devices\\[1em]
-Supplementary Material-\\[1em]}
\end{Large}

Daniel Rosenbach, Nico Oellers, Abdur R. Jalil, Martin Mikulics, Jonas Kölzer, Erik Zimmermann, Gregor Mussler, Stephany Bunte, Detlev Grützmacher, Hans Lüth and Thomas Schäpers\\[2em]

\begin{large}
\textbf{A. Temperature dependent magnetoconductance measurements on TLM nanoribbon segments of varying lengths}
\end{large}

\end{center}

\begin{figure*}[!htb]
    \centering
\includegraphics[keepaspectratio=true,width=\textwidth]{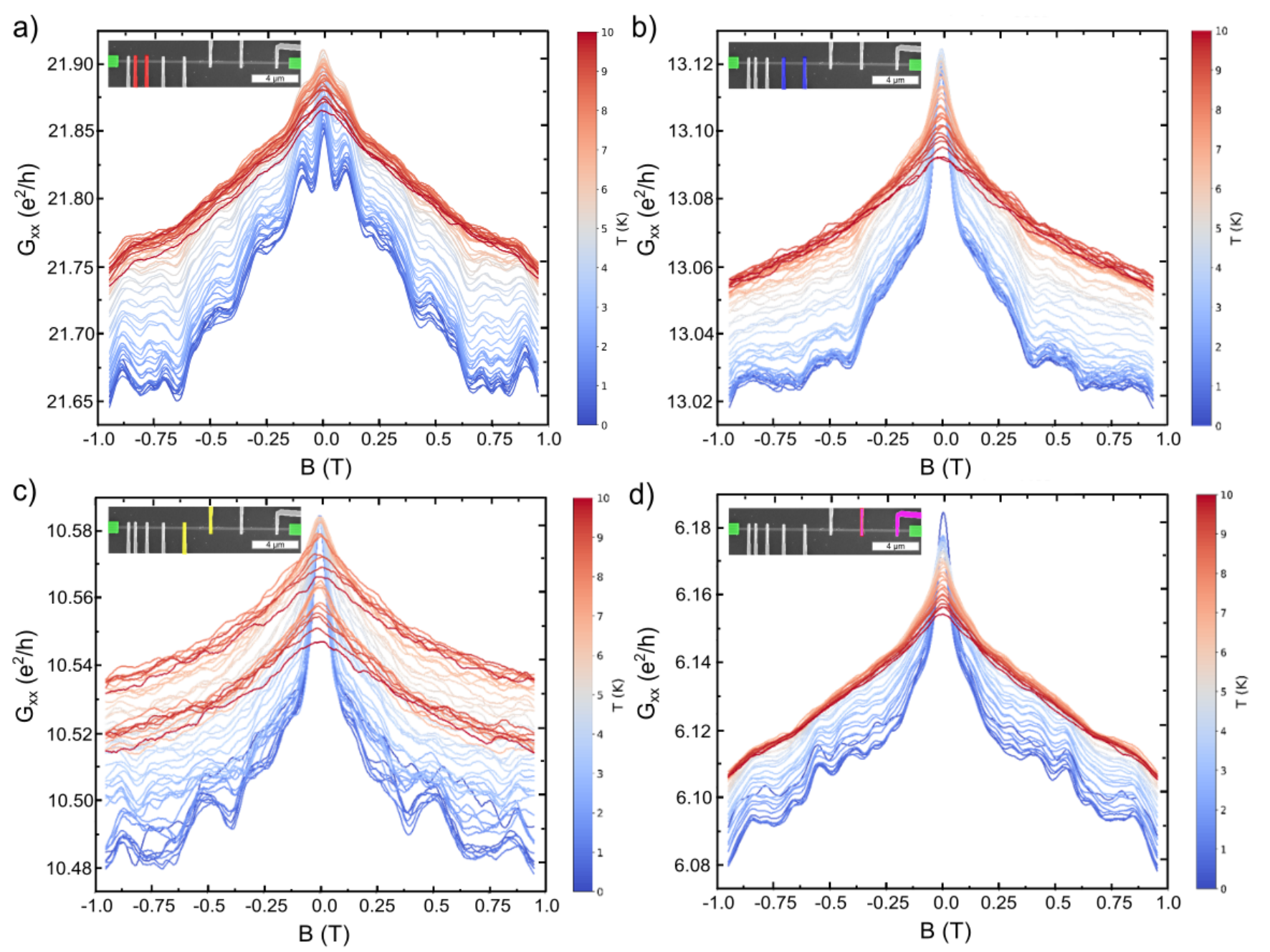}
    \caption{\textit{\textbf{Supplementary Fig. 1:}} Magnetoconductance measurements for different temperatures performed in a perpendicular applied magnetic field. Four different segments with a length $L$ of 950\,nm (a)), 1.9\,$\upmu$m (b)), 2.4\,$\upmu$m (c)) and 3.4\,$\upmu$m have been investigated. Temperatures range from 17\,mK (dark blue curves) to 10\,K (dark red curves). The color code for intermediate temperatures is given on the right, next to every figure.}
    \label{figS1}
\end{figure*}

\begin{figure*}[!htb]
	\centering
\includegraphics[keepaspectratio=true,width=\textwidth]{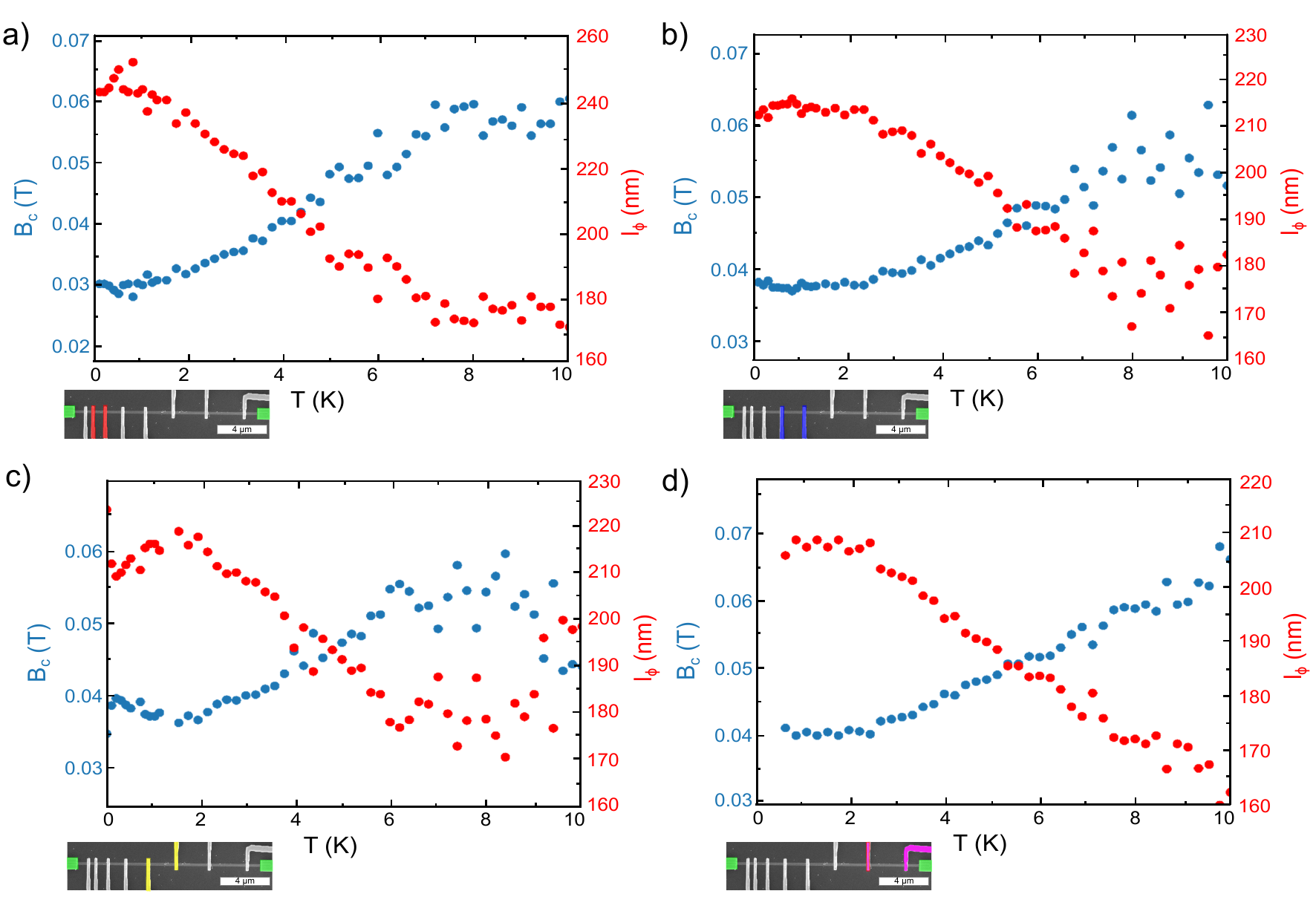}
    \caption{\textit{\textbf{Supplementary Fig. 2:}}Correlation field $B_c$ analysis of the temperature dependent magnetoconductance measurements in a perpedicular magnetic field. a)-d) show the correlation fields as well as phase coherence lengths determined for the different nanowire segments under consideration. The four terminal configurations for the measurements performed are shown in the inset of each figure. The green contact pairs have been used for current injection, other colors used display the pair of contacts in between which the potential difference has been measured. }
    \label{figS2}
\end{figure*}

The nanoribbon TLM device investigated in the main text consist of one single Bi$_2$Te$_3$ nanoribbon, which is separated into segments of different lengths, defined by the contact separation $L$ of two neighbouring Ti/Au contacts. The ribbon measures a constant width of around 200\,nm. Each and every segment of the nanoribbon can be considered a single nanoribbon device, where the contact seperation $L$ defines the device length. By applying a constant d.c. current along the whole nanoribbon, four-terminal measurements of the longitudinal conductance $G_{xx}$ as a function of the length $L$ of the different segments were performed.\\
Complimentary to the temperature dependent magnetoconductance in a perpendicular applied magnetic field shown in the main text Fig.~2 b) the same kind of measurements are shown on different segments of the nanoribbon device in Figs.~\ref{figS1} b) - d). In the inset of each figure, the corresponding pair of contacts is highlighted. As already pointed out in the main text, all curves show the same behavior at low temperatures. Sharp weak anti-localization (WAL) features were determined for every segment at 17\,mK, that are accompanied by specific universal conductance fluctuation (UCF) spectra. A fast fourier transformation (FFT) performed on the magnetoconductance data for all four segments is discussed in the main text. Additionally to the comparison of the FFT spectra at 17\,mK, the temperature dependency shows similar behavior for all four segments as well. This becomes more evident, when taking a look on the temperature dependent correlation fields $B_c$ as well as phase coherence lengths $l_{\phi}$ shown in Figs.~\ref{figS2}, determined using the correlation field analysis, as described in the maintext Eq. 2. The phase coherence length has been determined to increase from a value of $l_{\phi}=160\,$nm at 10\,K to a value of about$l_{\phi}=220\,$nm at about 2\,K. Below $T \sim 2\,K$ the phase coherence length does not increase anymore. In this range below 2\,K a correlation field of $B_c = 0.03-0.04\,$T has been determined for all four segments. The corresponding phase coherence length measures $l_{\phi}=210-240\,$nm.\\

\begin{center}
    
\begin{large}
\textbf{B. Nano Hall bar measurements}
\end{large}

\end{center}

\begin{figure*}[!htb]
	\centering
\includegraphics[keepaspectratio=true,width=\textwidth]{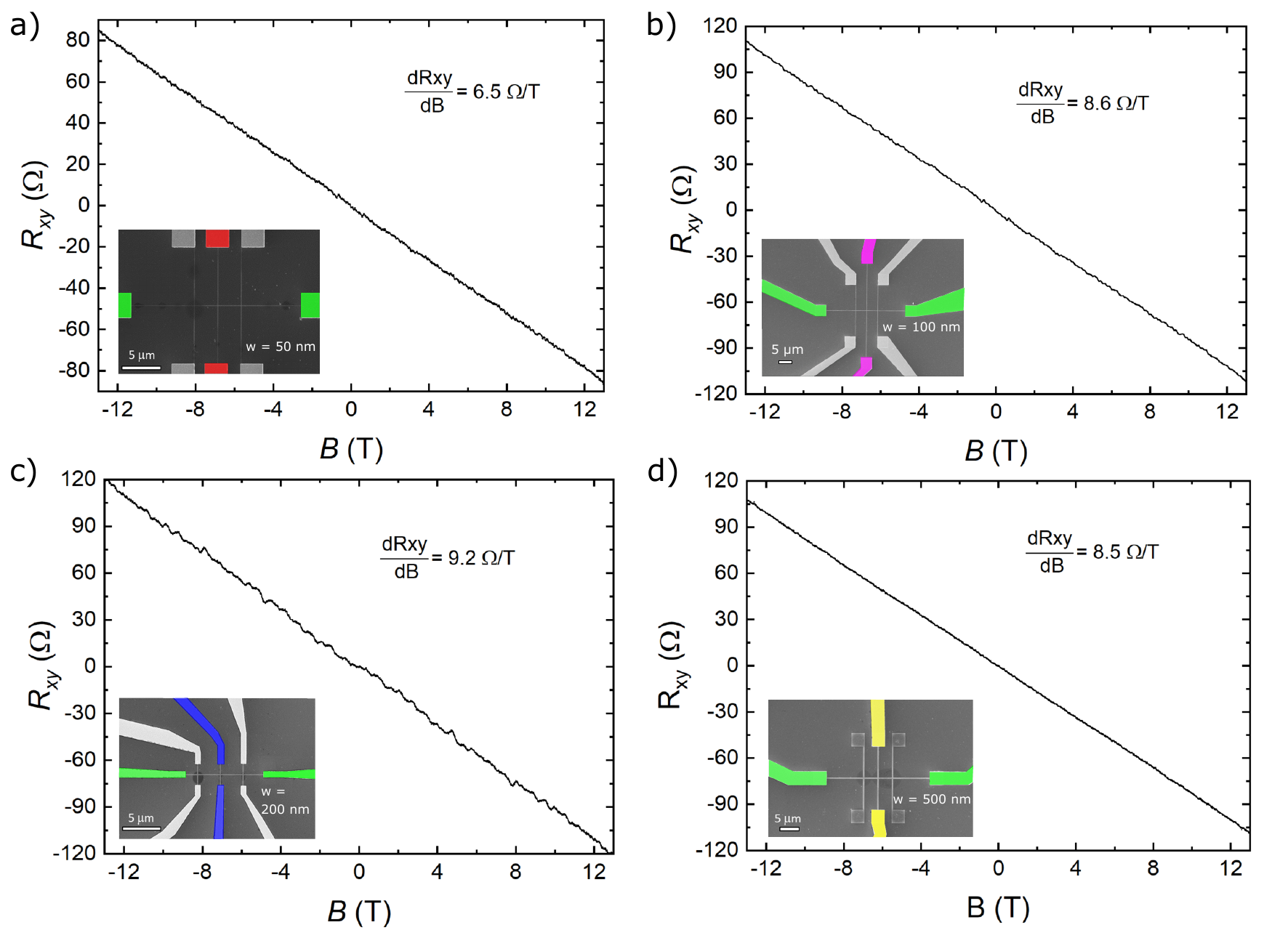}
    \caption{\textit{\textbf{Supplementary Fig. 3:}} Hall bar measurements performed on the a) 50\,nm, b) 100\,nm, c) 200\,nm and d) 500\,nm wide nanoribbons. In each inset the four terminal configuration for determination of the Hall resistance is displayed. The green contact pairs have been used for current injection, other colors used display the pair of contacts in between which the potential difference has been measured. Hall slopes $dR_{xy}/dB$ determined are displayed for every measurement, respectively.}
    \label{figS3}
\end{figure*}

Nano Hall bars of different channel widths $w$ have been investigated. An overview of the device dimensions is given in Tab. 1 in the main text. Along the nanoribbon an a.c. signal with an ampltiude of 10\,nA has been applied. Using a contact pair on opposite sites of the nanoribbon, perpendicular to the current flow, the Hall potential $U_{xy}$ has been measured. The Hall potential is given by

\begin{equation}
    U_{xy}=\frac{1}{ne}\frac{I}{w}B,
\end{equation}

where $n$ is the carrier concentration, $e$ the elemental charge, $I$ the applied bias and $w$ the nanoribbon width. The Hall resistance $R_{xy}=U_{xy}/I$ has been calculated and is displayed as a function of a perpendicular magnetic field $B$ in Figs.~\ref{figS3} a)-d). The sign of the Hall slope determines whether p-  or n-type conduction is present. For a mixed system the Hall curve would be non-linear. Following the device configuration and the direction of the magnetic field with respect to the current flow, all of the Hall bar measurements show n-type, electronic transport behavior. The Hall slope $dR_{xy}/dB$ has been determined for all measurements and is displayed in Figs.~\ref{figS3} a)-d). From this value, the two-dimensional sheet carrier concentration $n_{2D}=n\cdot w$ can be determined following

\begin{equation}
    n_{2D} = \frac{1}{q\cdot \frac{dR_{xy}}{dB}}.
\end{equation}

The sheet carrier concentrations lie in the range of $n_{2D}=(6.0-9.5) \times 10^{13}\,$cm$^{-2}$. From the sheet carrier concentration the mobility can be deduced using

\begin{equation}
    \upmu = \frac{1}{R_S\cdot en_{2D}},
\end{equation}
with $R_S$ being the sheet resistance. The mobility values determined lie in the range of $\upmu = (180-210)\,$cm$^2$/Vs.\\

\begin{figure*}[!htb]
	\centering
\includegraphics[keepaspectratio=true,width=\textwidth]{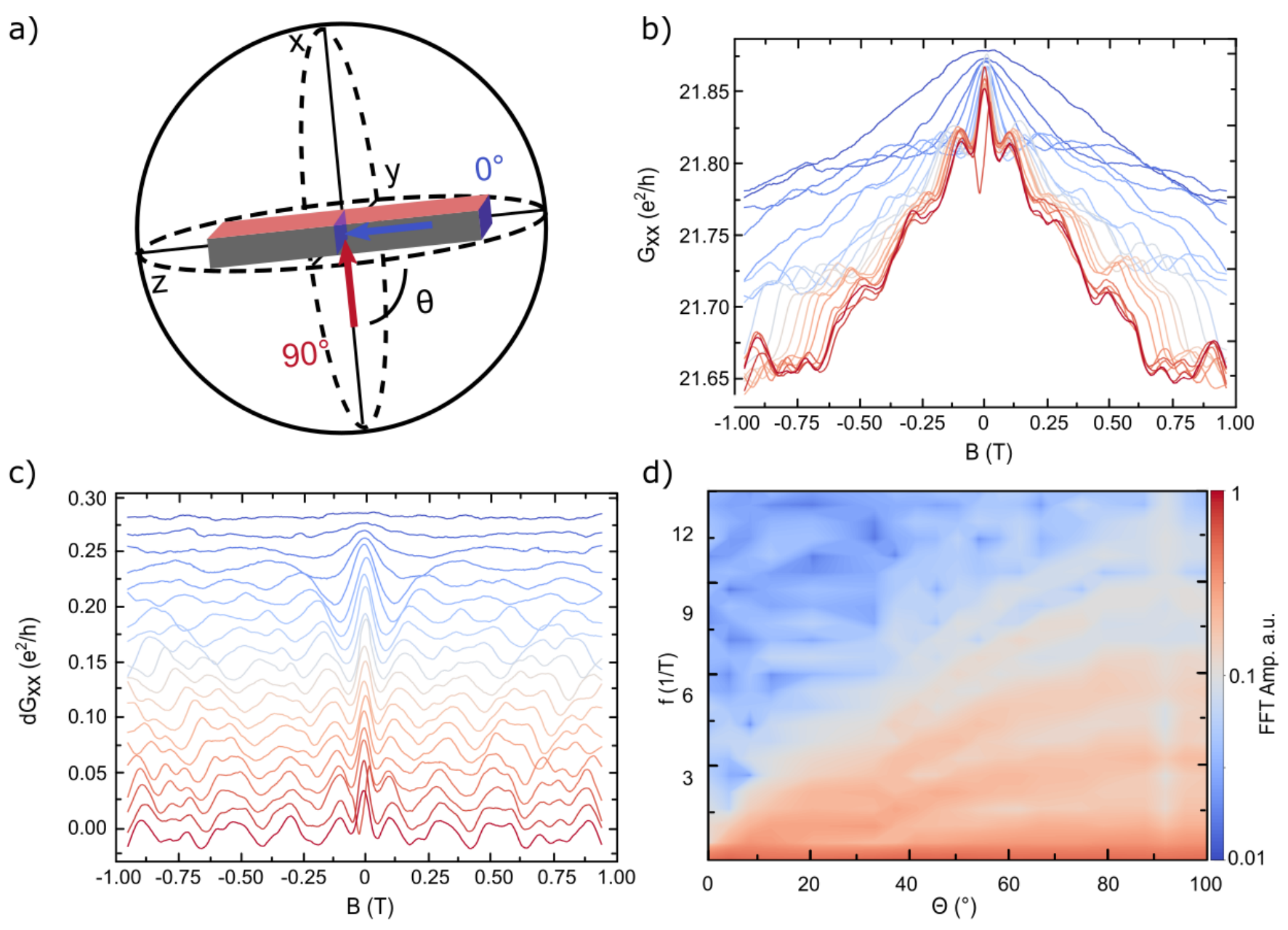}
    \caption{\textit{\textbf{Supplementary Fig. 4:}} Angle dependency of the magnetoconductance in between a perpendicular out-of plane and a parallel in-plane field orientation. a) Schematic representation of the applied magnetic field orientation in three dimensions. The angle $\Theta$ is changed from 0$^{\circ}$ (parallel in-plane) to 100$^{\circ}$. $\Theta = $90$^{\circ}$ corresponds to a perpendicular out-of plane orientation. b) Magnetoconductance curves for different angles taken at steps of $\Delta \Theta = 5\,^{\circ}$. c) Magnetoconductance at different angles after subtraction of a background using a first order SavitzkY--Golay filter. An constant offset is applied to make quantum modulations better visible. d) FFT amplitude as a function of angle and frequency. The FFT has been performed on the dataset in b).}
    \label{figS4}
\end{figure*}

\begin{figure*}[!htb]
	\centering
\includegraphics[keepaspectratio=true,width=\textwidth]{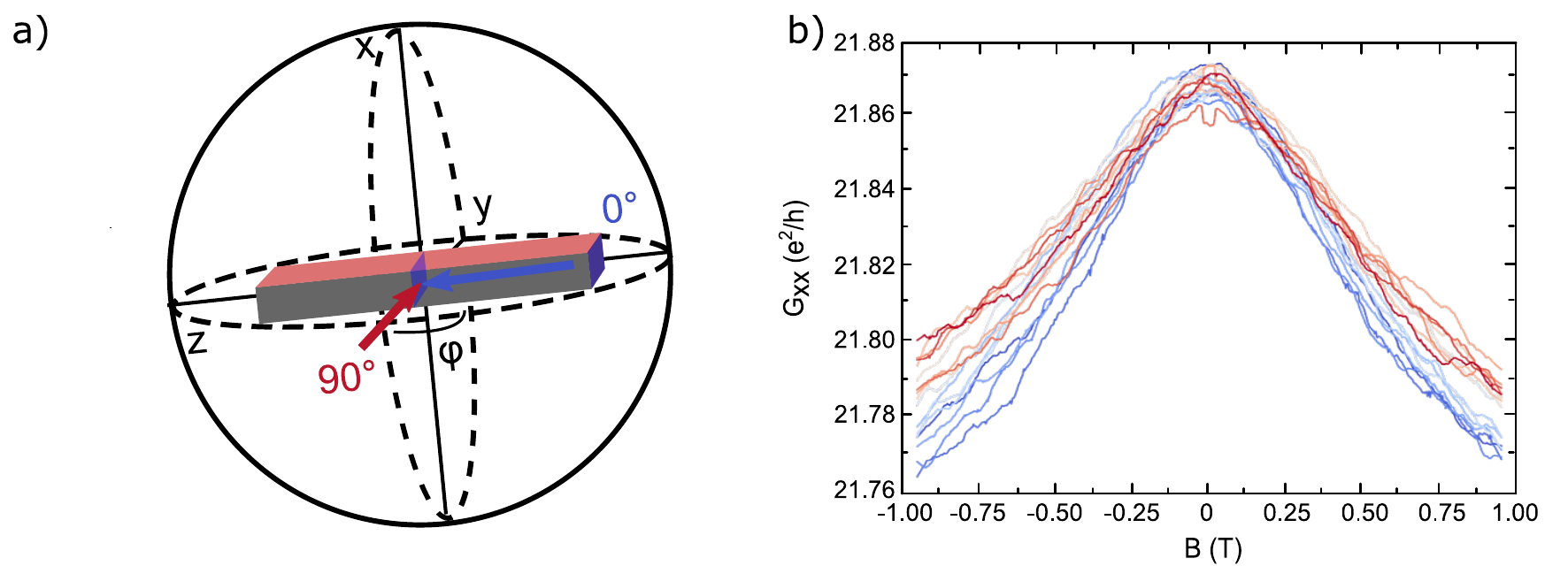}
    \caption{\textit{\textbf{Supplementary Fig. 5:}} Angle dependency of the magnetoconductance in between a perpendicular in-plane and a parallel in-plane field orientation. a) Schematic representation of the applied magnetic field orientation in three dimensions. The angle $\varphi$ is changed from 0$^{\circ}$ (parallel in-plane) to 100$^{\circ}$. $\Theta = $90$^{\circ}$ corresponds to a perpendicular in-plane orientation. b) Magnetoconductance curves for different angles taken at steps of $\Delta \Theta = 5\,^{\circ}$. }
    \label{figS5}
\end{figure*}

\begin{center}
    
\begin{large}
\textbf{C. Angle dependent magnetoconductance measurements on the TLM nanoribbon}
\end{large}

\end{center}
The TLM nanoribbon device has been measured in a dilution refrigerator equipped with a vector rotatable magnet. Three superconducting coils are installed. Up to $\pm$6\,T can be applied parallel to the nanoribbon, while the other two coils apply up to $\pm$1\,T in perpendicular in-plane and perpendicular out-of plane orientation. The position of the nanoribbon with respect to these three magnetic field axes is schematically shown in Figs. \ref{figS4} a) and \ref{figS5} a).\\
In a perpendicular out-of plane applied magnetic field (Fig.\ref{figS4} b), red curve) the magnetoconductance curve shows a pronounced WAL feature as well as UCFs, symmetrically around zero field. The angle dependence of these features has been discussed in the main paper for the nano Hall bars. The TLM nanoribbon structure behaves similarly in angle dependent measurements. Magnetoconductance measurements have been performed in a range of $0^{\circ} \leq \Theta \leq 100^{\circ}$ at a stepping of $\Delta \Theta = 5\,^{\circ}$. An angle of $\Theta = 90^{\circ}$ thereby corresponds to a magnetic field applied perpendicular out-of plane and an angle of $\Theta = 0^{\circ}$ to a field applied parallel in-plane.\\
The WAL feature changes, when the magnetic field is tilted. It appears that the WAL feature is broader in the parallel in-plane field orientation than it is in the perpendicular out-of plane orientation. The phase coherence length is directly related to the width of the WAL feature. For WAL features that span across a greater range of the magnetic field the phase-coherence length is smaller. When the magnetoconductance curves for different tilt angles are compared the phase-coherence length appears to be smallest in the in-plane parallel field orientation. A thorough discussion of the WAL fit with respect to closed loop interference in 3D TI nanoribbons is given in [1].\\
As has been discussed in the main paper the UCFs visible in Fig.\ref{figS4} b) appear to have different oscillation periods. These are increasing for angles $0^{\circ} \leq \Theta \leq 90^{\circ}$. Similarly to the UCFs discussed for the nano Hall bars, the effective area, perpendicular to the magnetic field axes, of loops parallel to the samples surface decreases as the magnetic field axis is gradually tilted from the perpendicular out-of plane into the parallel in-plane field orientation. This trend becomes better visible in the background subtracted magnetoconductance curves shown in Fig.\ref{figS4} c). The background has been subtracted by using a first order Savitzky--Golay filter. The single curves have a constant offset to make the quantum modulations better visible. The FFT performed on the dataset shown in Fig.\ref{figS4} b) is displayed in Fig.\ref{figS4} d). Single loops can be traced along the different tilt angles but the FFT lacks resolution, when compared to FFTs performed on the nano Hall bars in the main paper, due to the maximum magnetic field size of 1\,T.\\

The magnetic field has furthermore been tilted in between the parallel in-plane $\varphi = 0^{\circ}$ and the perpendicular in-plane $\varphi = 90^{\circ}$ field orientation, as schematically depicted in Fig.\ref{figS5} a). The field has thereby been tilted in steps of $\Delta \varphi = 5^{\circ}$. The recorded magnetoconductance curves are shown in Fig.\ref{figS5} b). It appears that the traces within the $\pm 1\,$T range do not differ much. The phase-coherence length appears to be of equal size for the parallel in-plane and the perpendicular in-plane field orientations.

\noindent
\underline{References:}\\

\noindent
[1] Kölzer, J., Rosenbach, D., Weyrich, C., Schmitt, T., Schleenvoigt, M., Jalil, A.R., Schüffelgen, P., Mussler, G., Sacksteder, V., Grützmacher, D., Lüth, H., and Schäpers, Th. Phase-coherent loops in selectively-grown topological insulator nanoribbons. \textit{arXiv e-prints}, \textbf{arXiv:} 1907.09801 (2019)